# COMPARATIVE STUDY OF SVD AND QRS IN CLOSED-LOOP BEAMFORMING SYSTEMS

Chau Yuen, Sumei Sun
Institute for Infocomm Research (I²R), Singapore
{cyuen, sunsm}@i2r.a-star.edu.sg

Jian-Kang Zhang
McMaster University, Canada
jkzhang@mail.ece.mcmaster.ca

*Abstract:* We compare two closed-loop beamforming algorithms, one based on singular value decomposition (SVD) and the other based on equal diagonal QR decomposition (QRS). SVD has the advantage of parallelizing the MIMO channel, but each of the sub-channels has different gain. QRS has the advantage of having equal diagonal value for the decomposed channel, but the subchannels are not fully parallelized, hence requiring successive interference cancellation or other techniques to perform decoding. We consider a closed-loop system where the feedback information is a unitary beamforming matrix. Due to the discrete and limited modulation set, SVD may have inferior performance to QRS when no modulation set selection is performed. However, if the selection of modulation set is performed optimally, we show that SVD can outperform QRS.

## I. INTRODUCTION

We consider a multiple-transmit and multiple-receive antennas wireless communications system (denoted as MIMO). Such system is well-known for achieving a higher capacity than traditional single-transmit single-receive antenna systems [1]. It has been adopted in many wireless communication standards, which include IEEE 802.11n for wireless local area network (LAN) application, 3GPP Long Term Evolution (LTE) for cellular communications.

Depending on the channel condition, the first generation MIMO technique aims at achieving a higher data rate, such as spatial multiplexing [2], or a higher diversity, such as space-time coding [3]. These techniques do not require the knowledge of channel state information (CSI) at the transmitter.

Due to the advancement in the communication technique, limited feedback information is possible in the future wireless communications system. It has been shown in [4] that the feedback of information on the CSI can greatly improve the system performance. In addition, closed-loop beamforming can also achieve lower decoding complexity, as we will show in the paper.

A straightforward beamforming algorithm is based on *singular value decomposition* (SVD) that fully diagonalizes the channel. However, each of the parallel sub-channels has different gain. In order to achieve better performance, different modulation can be applied to different stream of data, but this requires additional feedback information. Unfortunately, this solution is not optimal as the modulations are "discrete" (i.e. modulation size is in the form of power of two). Moreover, assigning the "appropriate" modulation for each of the data stream in real time is also a challenging task.

In this paper, we consider another beamforming algorithm based on *equal-diagonal QR decomposition* (QRS) from [5] (also known as *geometry mean decomposition*, GMD, in [6]). Unlike SVD, QRS does not fully diagonalize the channel, but converts the equivalent channel into an upper triangular matrix. Compared with SVD, QRS has the advantage of having equal diagonal elements for its equivalent channel. This implies that the same modulation can be applied to all the data streams. Hence it eliminates the trouble of assigning different modulation for different stream of data.

It has been shown in [7] that by using a simple and efficient closed-loop feedback, the performance of the MIMO-OFDM system for wireless LAN can be greatly improved. The extra cost is simply fed back the unitary precoders from SVD or QRS. However, no comparison between the SVD and QRS beamforming algorithms can be found in the literature. In this paper, we will provide a comparative study on SVD and QRS, with and without modulation set selection.

The organization of this paper is as follows: we first introduce the SVD and QRS decomposition in section II, the decomposition when the number of streams is less than the number of transmit and receive antennas for QRS scheme will also be discussed. Next, we will perform simulation to compare the two decomposition schemes in Section III. In the first comparison, we do not perform modulation set selection, while in the second comparison, we perform modulation set selection. We show that without modulation set selection, beamforming system based on QRS can outperform the one based on SVD, and vice versa if with modulation set selection. Finally we conclude the paper in Section IV.

## II. SVD AND QRS DECOMPOSITION

Consider a point-to-point MIMO system with $M$ transmit and $N$ receive antennas. The $N$–by–1 receive signal **y** can be modeled as follows:

$$\mathbf{y} = \mathbf{Hx} + \mathbf{n} \quad (1)$$

where **H** is the $N$–by–$M$ channel coefficient and **x** is the $M$–by–1 transmitted signal with **n** being the AWGN noise. In this paper, we consider Rayleigh fading channel, so **H** consists of zero-mean complex Gaussian random variables.

To achieve a better decoding performance and lower decoding complexity, beamforming can be applied. We first decompose the channel matrix **H** by SVD or QRS as follows:

$$\text{SVD:} \quad \mathbf{H} = \mathbf{UDV}^*$$
$$\text{QRS:} \quad \mathbf{H} = \mathbf{QRS}^* \quad (2)$$

where **U**, **V**, **Q**, **S** are all unitary matrices, **D** is a diagonal matrix with singular values as its elements, while **R** is an upper triangular matrix with identical diagonal elements. Superscript * denotes conjugate transpose.

The **D** and **R** are both of dimension $N$-by-$M$, and the rank, $d$, is limited by the minimum number of $M$ and $N$, i.e.

$$d \leq \min(M, N) \quad (3)$$

and for SVD, the diagonal matrix **D** consists of the singular values of the channel:

$$\mathbf{D} = diag(\delta_1, \delta_2, ..., \delta_d, 0, 0) \quad (4)$$

where $\delta_1, \delta_2, ... \delta_d$ are the singular values. While for QRS, the diagonal value, $r$, in **R** is the geometric mean of the singular value:

$$r = \left(\prod \delta_k\right)^{1/d}, \quad 1 \leq k \leq d \quad (5)$$

We consider the closed-loop beamforming system where only a unitary matrix, i.e. **V** in SVD and **S** in QRS, and the selection of the modulation set can be fedback from the receiver to the transmitter.

### A) SVD Beamforming

First consider SVD beamforming, the transmitted signal is:

$$\mathbf{x} = \mathbf{Vu} \quad (6)$$

where **V** is the unitary beamforming matrix obtained from SVD decomposition, and **u** is the intended data signal. The receiver of the SVD beamforming system can perform the following:

$$\mathbf{U}^*\mathbf{y} = \mathbf{U}^*(\mathbf{UDV}^*)\mathbf{Vu} + \mathbf{U}^*\mathbf{n}$$
$$= \mathbf{Du} + \tilde{\mathbf{n}} \quad (7)$$

Where the noise $\tilde{\mathbf{n}}$ is still AGWN as **U** is unitary. Hence simple decoding can be achieved, as **D** is a diagonal matrix. However, each of the sub-channel has a different gain in this case, as the diagonal value of **D** is related to the eigenvalue of the channel matrix **H**. As a result, to achieve optimal performance, different data streams should use a different modulation, however this would require extra feedback information, i.e., information on **D** needs to be fedback to the transmitter on top of **V**.

When there are fewer number of data streams than the maximum allowable, i.e. the minimum number of transmit and receive antennas, we can perform the SVD decomposition based on the first few principle eigenvectors. For example, in a four transmit four receive antenna system, if we are only required to transmit $n$ streams, where $n < d$, we can perform SVD with $\mathbf{V}_n$, where $\mathbf{V}_n$ consists of the first $n$ priniciple eigvenvectors of **H**.

### B) QRS Beamforming

Next consider QRS beamforming, similarly the transmitted signal is:

$$\mathbf{x} = \mathbf{Su} \quad (8)$$

where **S** is the unitary beamforming matrix obtained from QRS decomposition, and **u** is the intended data signal. The receiver of the QRS beamforming system can perform the following:

$$\mathbf{Q}^*\mathbf{y} = \mathbf{Q}^*(\mathbf{QRS}^*)\mathbf{Su} + \mathbf{Q}^*\mathbf{n}$$
$$= \mathbf{Ru} + \tilde{\mathbf{n}} \quad (9)$$

Since **R** is an upper triangular matrix, successive interference cancellation (SIC) or other algorithm can be performed at the receiver to decode data. The advantage of QRS is the diagonal elements of **R** are identical; this implies that we can apply the same modulation to all the data streams.

When we only need to transmit $n$ streams, where $n < d$, we can perform the QRS decomposition based on the required eigen subchannels, $\mathbf{H}_n$, where $\mathbf{H}_n$ is defined as:

$$\mathbf{H}_n = \mathbf{H}\mathbf{V}_n \quad (10)$$

and $\mathbf{V}_n$ consists of the first $n$ priniciple eigvenvectors of $\mathbf{H}$.

### III. SIMULATION RESULTS

We consider a Rayleigh flat fading uncoded MIMO system with four transmit and four receive antennas. In addition, throughout the simulation, it is assumed that the channel is estimated correctly, and the beamforming matrix is perfectly known at the transmitter without any delay. SIC will be employed for system with QRS-based beamforming. The transmission is fixed at spectral efficiency of 8 bps/Hz, which may be realized with two, three or four data streams. The available modulation sets for SVD and QRS are as follows:

*Modulation for SVD with 8bps/Hz:*

Two streams: {QAM64-QPSK}

Two streams: {QAM16-QAM16}

Three streams: {QAM16-QPSK-QPSK}

Three streams: {QAM8-QAM8-QPSK}

*Modulation for QRS with 8bps/Hz:*

Two streams: {QAM16-QAM16}

Three streams: {QAM16-QPSK-QPSK}

Three streams: {QAM8-QAM8-QPSK}

Four streams: {QPSK-QPSK-QPSK-QPSK}

Due to the uneven nature on the diagonal values of the effective channel on SVD, we have a two streams uneven modulation QAM64-QPSK for SVD. In contrast, due to the equal value on the diagonal values of the effective channel on QRS, we have a four streams equal modulation QPSK for QRS.

We compare the BER performance in Figure 1 when the selection of modulation set is not allowed, i.e. same modulation is used throughout the simulation. For SVD, modulation set with {16QAM-16QAM} and {16QAM-QPSK-QPSK} give the best BER performance. For QRS, modulation set with {16QAM-16QAM} give the best BER performance. So from Figure 1, we can conclude that without the modulation set selection, QRS greatly outperforms SVD at high SNR region, about 1dB at BER of $10^{-3}$.

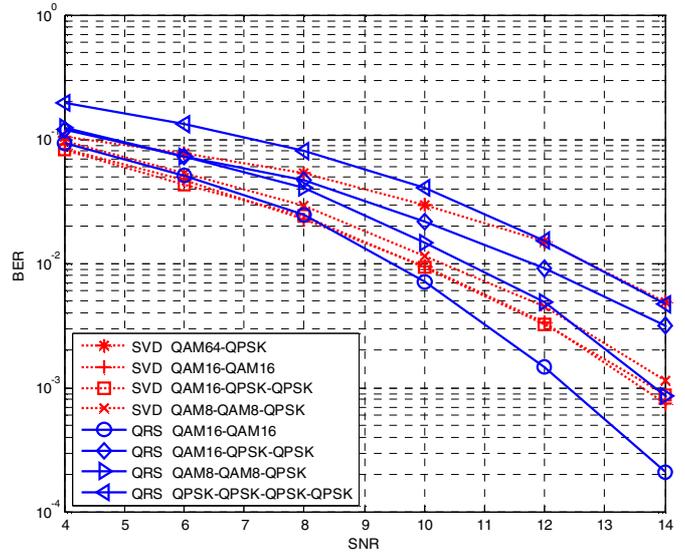

Figure 1 Simulated BER for 4tx-4rx at 8bps/Hz without modulation set selection

We then compare BER performance in Figure 2 when the selection of modulation set is allowed. The selection is optimal, as the selection is performed "offline", i.e. we pick the modulation set that gives the best BER. Hence this serves as a lower bound for the actual BER performance with "online" modulation set selection.

From Figure 2, we observe that by allowing the selection of modulation set, SVD gives a better overall performance, about 1dB at BER of $10^{-3}$. The gap between the selection of modulation set versus the fixed single modulation set {QAM16-QAM16} is 4.5dB for SVD at BER $10^{-3}$, while the gap is 2.5dB for QRS at BER $10^{-3}$. This suggests that the performance of SVD relies heavily on the selection of the modulation set, which requires additional feedback information. In addition, it still remains unclear on the optimal way to select modulation set in a practical MIMO OFDM system when it involves a large number of sub-carriers. The gain that is demonstrated in Figure 2 is based on optimal selection, the actual gain would be reduced due to non-optimal selection and delay in feedback.

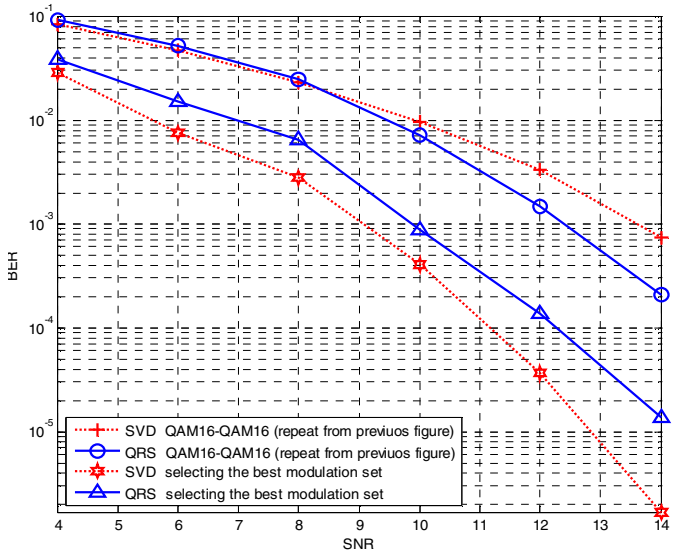

Figure 2 Simulated BER for 4tx-4rx at 8bps/Hz with modulation set selection

For SVD, a technique called "power loading" can help to mitigate the disadvantages of discrete modulation, but it requires additional computational complexity and feedback overhead.

## IV. CONCLUSION

We have shown that closed-loop MIMO system with Equal-Diagonal QR decomposition (QRS) may outperform the one based on SVD decomposition. This is due to fact that there is only a limited choice of modulation sets, and the available modulation sets are discrete, hence getting the optimal modulation set matched to the channel gain provided by SVD may not be always possible. However, when the selection of modulation set is allowed, SVD may achieve a better performance. We further show that the performance of SVD relies heavily on the selection of the appropriate modulation set.

In [8], pseudo-inverse is proposed to achieve better decoding performance. It would be interesting to look at the channel decomposition of the equivalent pseudo-inverse channel, for both SVD and QRS. The coded performance of these two systems will be an interesting topic for future work.